\documentclass[fleqn,11pt]{wlscirep}
\usepackage[utf8]{inputenc}
\usepackage[T1]{fontenc}
\usepackage{longtable}
\title{Examining the Global Spread of COVID-19 Misinformation}

\author[1]{Sophie Nightingale}
\author[2,*]{Hany Farid}
\affil[1]{Department of Psychology, Lancaster University, Lancaster, UK}
\affil[2]{Electrical Engineering \& Computer Sciences and School of Information, University of California, Berkeley, Berkeley CA, USA}

\affil[*]{hfarid@berkeley.edu}

\keywords{COVID-19 $|$ coronavirus $|$ misinformation $|$ conspiracies $|$ social media} 

\begin{abstract}
The global COVID-19 pandemic has led to the online proliferation of health-, political-, and conspiratorial-based misinformation. Understanding the reach and belief in this misinformation is vital to managing this crisis, as well as future crises. The results from our global survey finds a troubling reach of and belief in COVID-related misinformation, as well as a correlation with those that primarily consume news from social media, and, in the United States, a strong correlation with political leaning.
\end{abstract}
\begin{document}

\flushbottom
\maketitle
\thispagestyle{empty}

\section*{Introduction}

The COVID-19 global pandemic has been an ideal breeding ground for online misinformation and conspiracies: Social-media traffic has reached an all-time record~\cite{facebook_newsroom} as people are forced to remain at home, often idle, anxious, and hungry for information~\cite{pakpour2020}, while at the same time, social-media services are unable to rely on human moderators to enforce their rules~\cite{washington_post}. The resulting spike in COVID-19 related misinformation is of grave concern to health professionals~\cite{avaaz}. 

A vaccine holds the potential to save untold lives; yet since July 2020 there has been significant growth in anti-vaccine misinformation on social media  with accounts of ``anti-vaxxers" gaining at least $7.8$ million followers since 2019~\cite{CCDH2020}. The consequences of this movement include a general erosion of trust in scientific expertise as well as opposition to vaccination --- between May and September of this year, intent to get vaccinated has declined in the United States~\cite{malik2020, pewresearch}. Based on the growing momentum of the anti-vaxx movement, researchers predict that, in a decade, such anti-vaccination views will be dominant~\cite{johnson2020}. As a result of this landscape, the World Health Organization has listed the need for surveys and qualitative research about the {\em infodemic} in its top priorities to contain the pandemic~\cite{WHO2020}.

A recent survey confirmed that belief in COVID-19 conspiracy theories is associated with smaller compliance with public health directives~\cite{allington2020}. Another recent study found that political affiliation is a strong predictor of knowledge of COVID-19 related information~\cite{nielsen2020}. Building on this earlier work, we launched a large-scale, multi-lingual, global study to examine the belief in $21$ prevalent COVID-19 related false statements, and $21$ corresponding true statements, Table~\ref{tab:datasetS1}~\footnote{All data will be made available in a public data repository}. We evaluate the reach and belief in these statements and correlate the results with political leaning, primary source of media consumption, and intent to vaccinate.


\begin{figure}[t]
    \centering
    \includegraphics[width=12cm]{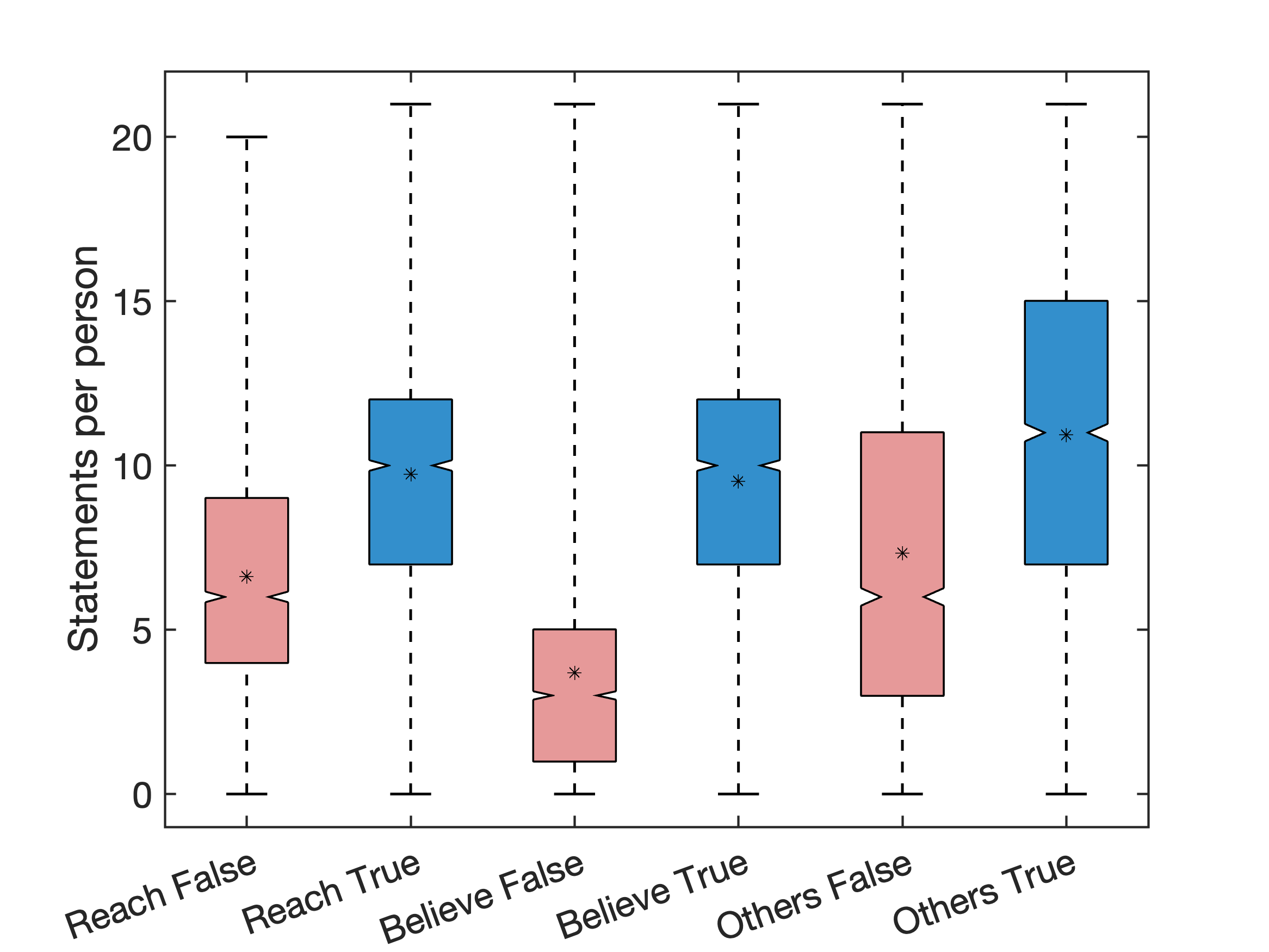}
    \caption{Averaged across all four geographic regions (N = $2,095$), the number of false and true statements, out of $21$ each, that reached respondent (they directly heard or read it), are believed by respondent, and are believed by someone known to respondent. The black asterisks correspond to the mean, the kink in the box plot corresponds to the median, the box plot height corresponds to the interquartile range ($1^{\mbox{st}}-3^{\mbox{rd}}$ quartile), and the dashed lines correspond to the minimum and maximum.}
    \label{fig:overallRBO}
\end{figure}


\section*{Methods}

A total of $2,708$ respondents were recruited from Ipsos’ online panel.\footnote{This research was approved by UC Berkeley's Office for Protection of Human Subjects, Protocol ID: 2019-08-12441. Respondents gave informed consent prior to taking part.} Ipsos is a global company that specializes in market research and public opinion. Respondents includes a diverse sample across more than $100$ countries. To ensure appropriate representation, Ipsos uses quota sampling to provide a sample that is matched to country demographics based on age, gender, ethnicity, and geography. respondents were recruited from countries spanning four regions of the world: Central and South America (CSA), Middle East and North Africa (MENA), United States (US), and Western Europe (WE) (for more details about the population and sample demographics see Table~\ref{tab:datasetS2} in the Supplementary Materials). Ipsos reported a $75\%$ response rate corresponding to an AAPOR response rate of 1.

Respondents were instructed they would participate in a study evaluating COVID-19 related misinformation. They read, one at a time, $42$ statements, Table~\ref{tab:datasetS1}, half of which are true and half are not, and specify: (1) if they had seen/heard the statement before; (2) if they believed the statement to be true; and (3) if they know someone that believes or is likely to believe the statement. With self-censorship of personal views on the rise~\cite{Ekins20}, asking participants what they think others believe might provide a more accurate indication of regional views~\cite{Galesic20}. The study was available in six languages (Arabic, English, French, German, Italian, Spanish), and respondents selected their preferred language. 

The $42$ statements were sourced from reputable fact-checking websites. To ensure a balanced design, each false statement was matched with a similarly themed true statement. Of the originally selected statements, four of the $21$ false statements and seven of the $21$ true statements contained content specific to the United States, and were only shown in this region. For the other three regions, each of these US-specific statements was replaced with a statement consistent in theme but with content that was considered more relevant in these other regions. The $42$ statements plus three attention-check questions, Table~\ref{tab:datasetS1}, were presented in a random order. After responding to all statements, respondents were asked six questions relating to the COVID-19 crisis: their intention to get vaccinated, perceptions of their government's handling of the crisis, whether, globally, they believe that the situation is getting better or worse, their own level of fear about getting the virus, and their general level of happiness. At the end of the survey, respondents were also asked how they consume news (television, newspapers (print or online), radio, social media, word of mouth), their political leaning (strongly left, slightly left, center, slightly right, strongly right), and basic demographics: education-level, age, gender, and race (U.S. only). All responses were collected between July 14, 2020 and July 28, 2020, amidst the global COVID-19 crisis.

The three, obviously false, attention-check questions were used to ensure that respondents were paying attention to the survey. A respondent's data was discarded if they failed to correctly answer any of these attention-check questions: in CSA $100$ of the $652$ responses were discarded, yielding $552$ usable responses; in MENA $160$ of the $676$ responses were discarded, yielding $516$ responses; in the US, $203$ of the $721$ responses were discarded, yielding $518$ responses; and in WE, $150$ of the $659$ responses were discarded, yielding $509$  responses. At the end of the study, respondents were again informed that half of the statements they read were not true, asked to confirm that they understood this, and were directed to several websites with accurate health information.
 

\section*{Results}

\subsection*{Central and South America (CSA)}

On average, $52.6\%$~|~$38.9\%$ of true~|~false statements, Table~\ref{tab:datasetS1}, reached respondents, $51.0\%$~|~$21.8\%$ of true~|~false statements are believed, and $58.7\%$~|~$44.0\%$ of true~|~false statements are believed by others known to the respondent. The median number of true~|~false statements that reached a respondent is $11$~|~$8$, the median number of true~|~false statements believed by a respondent is $11$~|~$4$, the median number of true~|~false statements believed by others known to the respondent is $12$~|~$9$, and $77.4\%$ believe at least one false conspiracy (false statements 10-14 in Table~\ref{tab:datasetS1}).

The results of six negative binomial regression models reveal only small effects of respondent demographics on the reach, belief, and belief by others in the true~|~false statements. The predictor variables are respondent gender, age, education, political leaning, and main news source. Political leaning did not predict the likelihood of believing any of the statements, and main news source influenced the likelihood of believing only $1~|~21$ false statements. 


\begin{table*}[p]
\resizebox{\textwidth}{!}{%
\def\arraystretch{0.8}%
\begin{tabular}{c|p{18cm}|c@{\hspace{0.5em}}c@{\hspace{0.5em}}c|c@{\hspace{0.5em}}c@{\hspace{0.5em}}c|c@{\hspace{0.5em}}c@{\hspace{0.5em}}c|c@{\hspace{0.5em}}c@{\hspace{0.5em}}c}
    \hline
         & &  \multicolumn{3}{|c|}{US (\%)} & \multicolumn{3}{|c|}{WE (\%)} & \multicolumn{3}{|c|}{MENA (\%)} & \multicolumn{3}{|c}{CSA (\%)} \\
         & STATEMENT (FALSE) & R & B & O & R & B & O & R & B  & O & R & B & O \\
         \hline
        1 & Gargling warm water with salt or vinegar eliminates infection from COVID-19 & 18 & 8 & 29 & 21 & 3 & 20 & 59 & 21 & 52 & 61 & 14 & 55 \\
        2 & Drinking sodium bicarbonate and lemon juice reduces the acidity of the body and the risk of getting infected with COVID-19 & 13 & 6 & 27 & 15 & 4 & 18 & 40 & 19 & 36 & 58	&	22	&	54 \\
        3 & If you can take a deep breath and hold it for more than 10 seconds without coughing and without difficulty it shows that you do not have the COVID-19 infection & 29	&	10	&	34	&	33	&	7	&	26	&	56	&	27	&	53	&	47	&	22	&	50 \\
        4 & Hand sanitizer has the same ingredients as antifreeze and can be fatal to pets if they lick your hand & 11	&	16	&	32	&	8	&	15	&	25	&	14	&	17	&	27	&	14	&	15	&	32 \\
        5 & Breathing in hot air from a hair dryer or in a sauna can prevent or cure COVID-19 & 13	&	4	&	20	&	12	&	3	&	12	&	23	&	7	&	28	&	25	&	6	&	33 \\
        6 & Taking large doses of Vitamin C will protect you from COVID-19 & 32	&	14	&	36	&	39	&	8	&	28	&	69	&	32	&	57	&	72	&	30	&	68 \\
        7 & Silver solution kills COVID-19 & 9	&	4	&	19	&	7	&	3	&	12	&	9	&	5	&	18	&	14	&	5	&	23 \\
        8 & Drinking or gargling with bleach can cure COVID-19 & 38	&	3	&	18	&	34	&	1	&	10	&	27	&	4	&	19	&	32	&	2	&	26 \\
        9 & The COVID-19 outbreak is not actually caused by a virus, but by 5G-wireless phone technology & 39	&	5	&	23	&	48	&	3	&	27	&	43	&	4	&	26	&	51	&	4	&	35 \\
        10 & Asymptomatic carriers of COVID-19 (infected but no symptoms) who die of other medical problems are added to the coronavirus death toll to get the numbers up to justify this pandemic response & 49	&	37	&	52	&	42	&	33	&	43	&	52	&	41	&	51	&	69	&	55	&	71 \\
        11 & No one is sick with COVID-19, the media is instilling “irrational fear” & 45	&	9	&	39	&	39	&	4	&	23	&	54	&	8	&	41	&	61	&	6	&	45 \\
        12 & A vaccine for COVID-19 has been used on cows for years & 6	&	5	&	19	&	4	&	4	&	12	&	8	&	6	&	15	&	7	&	7	&	23 \\
        13 & COVID-19 was man-made in a lab and is not thought to be a natural virus & 71	&	26	&	50	&	81	&	30	&	56	&	91	&	43	&	73	&	91	&	51	&	82 \\
        14 & Bill Gates plans to use COVID-19 to implement a mandatory vaccine program with microchips to track people & 44	&	22	&	43	&	38	&	13	&	34	&	46	&	25	&	46	&	53	&	23	&	54 \\
        15 & House Democrats included \$25 million to boost their own salaries in their proposal for the COVID-19 related stimulus package (US) / British MPs awarded themselves a hand-out of £10,000 to work from home during the COVID-19 crisis (WE, MENA, CSA)  & 21	&	32	&	50	&	9	&	28	&	33	&	15	&	32	&	36	&	16	&	34	&	39 \\
        16 & COVID stands for ``Chinese Originated Viral Infectious Disease" & 17	&	16	&	40	&	13	&	11	&	24	&	30	&	25	&	37	&	22	&	18	&	40 \\
        17 & US President Donald Trump tweeted that economic stimulus checks would only go to those who have not made negative social media comments about him (US) / In a speech, President Vladimir Putin blamed the West for the spread of COVID-19 (WE, MENA, and CSA) & 10	&	13	&	34	&	23	&	26	&	35	&	25	&	30	&	42	&	32	&	40	&	53 \\
        18 & Multiple states have banned alcohol sales due to COVID-19 (US) / UK’s National Health Service (NHS) banned alcohol consumption and sales in an effort to combat COVID-19 (WE, MENA, CSA) & 21	&	22	&	34	&	12	&	11	&	18	&	19	&	22	&	32	&	22	&	35	&	40 \\
        19 & Russia unleashed more than 500 lions on its streets to ensure that people are staying indoors during this pandemic outbreak & 14	&	7	&	21	&	10	&	4	&	12	&	35	&	14	&	28	&	36	&	15	&	31 \\
        20 & Staff at Gold Coast Hospital in Australia rolled in a Wilson volleyball to keep Tom Hanks company in quarantine after he was diagnosed with COVID-19 & 15	&	29	&	36	&	9	&	23	&	28	&	19	&	31	&	34	&	15	&	33	&	38 \\
        21 & Sales of Corona beer dropped sharply in early 2020 because consumers mistakenly associated the brand name with the new coronavirus (US) / Making the most of humans being in social isolation, a group of elephants broke into a village in Yunan province and got drunk on corn wine (WE, MENA, CSA) & 53	&	48	&	52	&	8	&	13	&	18	&	15	&	16	&	26	&	21	&	20	&	33 \\
        \hline 
        & STATEMENT (TRUE) \\
        \hline 
        1 & Rubbing alcohol products that are at least 70 percent alcohol will kill COVID-19 & 49 & 41 & 52 & 50 & 36 & 48 & 57 & 42 & 57 & 83 & 64 & 79 \\
        2 & Hundreds of people have died in Iran after consuming methanol in an effort to treat COVID-19 & 15 & 25 & 33 & 17 & 28 & 31 & 32 & 36 & 44 & 20 & 36 & 42 \\
        3 & Home-made masks do not stop you from getting COVID-19, but can reduce the risk of spreading the virus if you already have it & 70	&	58	&	65	&	79	&	70	&	71	&	83	&	74	&	76	&	88	&	75	&	83 \\
        4 & The virus causing COVID-19 can survive for multiple days on solid surfaces like plastic or stainless steel & 71	&	52	&	66	&	75	&	62	&	69	&	86	&	66	&	75	&	91	&	76	&	86 \\
        5 & There is currently no treatment or cure for COVID-19 & 78	&	61	&	69	&	82	&	68	&	71	&	88	&	70	&	78	&	89	&	64	&	79 \\
        6 & Trump touts hydroxychloroquine, an anti-malarial drug, as a cure for COVID-19 (US) / There is currently no evidence that pets can transmit COVID-19 to people (WE, MENA, CSA) & 70	&	25	&	53	&	68	&	50	&	55	&	74	&	52	&	59	&	81	&	61	&	67 \\
        7 & COVID-19 is spread through respiratory droplets, which typically travel about three to six feet and settle on surfaces & 81	&	75	&	76	&	84	&	77	&	80	&	84	&	78	&	78	&	87	&	82	&	85 \\
        8 & Household bleach is an effective household cleaner for COVID-19 & 66	&	55	&	62	&	47	&	32	&	42	&	57	&	40	&	54	&	75	&	60	&	77 \\
        9 & In the UK, at least a dozen 5G-wireless phone towers have been vandalized amidst claims linking COVID-19 to 5G networks & 20	&	30	&	36	&	34	&	35	&	38	&	30	&	25	&	37	&	38	&	33	&	42 \\
        10 & In February, the White House chief of staff accused the US media of stoking a coronavirus panic in the hope it would take down President Donald Trump (US) / Research suggests that the earliest cases of COVID-19 in New York originated with travelers coming from Europe, not Asia (WE, MENA, CSA) & 46	&	33	&	48	&	28	&	27	&	36	&	39	&	37	&	48	&	63	&	63	&	66 \\
        11 & On March 15, Rodney Howard-Browne, the pastor of a Pentecostal megachurch in Florida, told his church to greet each other with a handshake (US) / More than 300,000 people worldwide have died from COVID-19 related illnesses (WE, MENA, CSA) & 21	&	41	&	45	&	76	&	78	&	76	&	88	&	80	&	87	&	93	&	84	&	88 \\
        12 & The US facilitated the sending of nearly 17.8 tons of donated medical supplies to China to combat the spread of COVID-19 in early 2020 & 31	&	36	&	43	&	19	&	25	&	31	&	28	&	34	&	42	&	31	&	35	&	45 \\
        13 & A train engineer tried to drive a freight train into USNS Mercy, the Navy medical ship providing relief to hospitals overburdened with COVID-19 patients (US) / The coronavirus pandemic has led to an unprecedented global surge in digital surveillance (WE, MENA, CSA) & 16	&	21	&	29	&	43	&	48	&	50	&	62	&	63	&	62	&	63	&	72	&	67 \\
        14 & A Bill Gates funded UK-based lab has a patent on a strain of coronavirus that affects only chickens & 7	&	9	&	25	&	7	&	8	&	18	&	12	&	15	&	25	&	15	&	14	&	30 \\
        15 & US President Donald Trump golfed several times and held a number of rallies after learning about the threat of COVID-19 (US) / In early April, both Germany and France accused the US of diverting scarce medical supplies meant for their respective countries to the US (WE, MENA, CSA) & 63	&	66	&	66	&	42	&	46	&	52	&	37	&	37	&	45	&	34	&	38	&	46 \\
        16 & At a speech in March 2020 US President Donald Trump's notes had the word ``Corona" scratched out and replaced in handwriting by ``Chinese." & 31	&	41	&	49	&	21	&	34	&	41	&	31	&	34	&	42	&	25	&	36	&	45 \\
        17 & In the US, the week ending 28th March 2020 saw a record 6.65 million initial unemployment claims filed (US) / Britain missed three opportunities to be part of an EU scheme to bulk-buy masks, gowns and gloves to protect against COVID-19 (WE, MENA, CSA) & 64	&	72	&	73	&	16	&	27	&	30	&	18	&	30	&	33	&	12	&	29	&	32 \\
        18 & Some US police departments have put out fake warnings that illicit drugs could be contaminated COVID-19 (US) / The United Nations’ biodiversity chief has called for a global ban on wildlife markets – such as the one in Wuhan, China (WE, MENA, CSA) & 9	&	14	&	29	&	40	&	55	&	50	&	43	&	50	&	50	&	55	&	62	&	63 \\
        19 & A Bronx zoo tiger has tested positive for COVID-19 & 51	&	46	&	53	&	33	&	36	&	43	&	24	&	28	&	36	&	38	&	36	&	47 \\
        20 & Actor Matthew McConaughey hosted a bingo night for the elderly isolating Texans & 12	&	31	&	36	&	7	&	28	&	25	&	8	&	28	&	29	&	7	&	21	&	28 \\
        21 & A man ran a marathon on his 23-foot balcony during COVID-19 lockdown & 16	&	33	&	35	&	28	&	42	&	42	&	15	&	31	&	35	&	18	&	31	&	35 \\
        \hline
                 & STATEMENT (ATTENTION) \\
        \hline
        1 & In December 2019, a spaceship of aliens came to Earth and killed millions of humans &-&-&-&-&-&-&-&-&-&-&-&- \\
        2 & A new species of tree that grows money has been discovered by tourists holidaying in the Amazon rain forest &-&-&-&-&-&-&-&-&-&-&-&-  \\
        3 & Scientists at the University of Oxbridge have discovered that humans can survive and live a normal life without either a brain or a heart  &-&-&-&-&-&-&-&-&-&-&-&-
\end{tabular}
}
\caption{Twenty-one false (top) and true (middle) statements and three obviously false attention-check statements. Shown in the four right most columns is, per statement, the percentage of statements that reached (R), are believed by (B), and are believed by others (O) in each geographical region.}
\label{tab:datasetS1}
\end{table*}


\subsection*{Middle East and North Africa (MENA)}

On average, $47.4\%$~|~$35.6\%$ of true~|~false statements reached respondents, $45.2\%$~|~$20.5\%$ of true~|~false statements are believed, and $52.0\%$~|~$37.0\%$ of the true~|~false statements are believed by others known to the respondent. The median number of true~|~false statements that reached a respondent is $10$~|~$7$, the median number of true~|~false statements believed by a respondent is $10$~|~$4$, the median number of true~|~false statements believed by others known to the respondent is $11$~|~$8$, and $65.9\%$ believe at least one false conspiracy.

The results of six negative binomial regression models reveal only small effects of respondent demographics on the reach, belief, and belief by others in the true~|~false statements. Political leaning did not predict  the likelihood of believing any of the statements, and main news source influenced the likelihood of believing only $1~|~21$ false statements. 


\subsection*{United States (US)}

On average, $42.3\%$~|~$27.1\%$ of true~|~false statements reached respondents, $41.2\%$~|~$16.0\%$ of true~|~false statements are believed, and $49.6\%$~|~$33.7\%$ of the true~|~false statements are believed by others known to the respondent. The median number of true~|~false statements that reached a respondent is $9$~|~$6$, the median number of true~|~false statements believed by a respondent is $9$~|~$3$, the median number of true~|~false statements believed by others known to the respondent is $10$~|~$6$, and $54.3\%$ believe at least one false conspiracy.

We conducted six negative binomial regression models with six outcome variables corresponding to the reach, belief, and belief by others in each true~|~false statement. Political leaning and main news source had an effect on the likelihood of the number of false statements believed. The number of false statements believed by those on the right of the political spectrum\footnote{From $518$ responses, $125~|~137$ reported as politically left~|~right of center, and $256$ as center. For the evaluation of the impact of political leaning, only those reporting left~|~right of center were considered.} is $1.91$ times greater than those on the left ($95\%$~CI [$1.54$, $2.37$]). The number of false statements believed by those with social media as their main source of news is $1.40$ times greater than those who cited another main news source ($95\%$~CI [$1.19$, $1.66$]).

We next performed a binary logistic regression to evaluate how political leaning and main news source influenced belief in each false statement. Political leaning influenced the likelihood of believing $12~|~21$ false statements, and main news source influenced the likelihood of believing $9~|~21$ false statements. For $11~|~12$ false statements where there is an effect of political leaning, those on the right are more likely to believe the false information. For all $9$ false statements, where there was an effect of main news source, those with social media as a main source are more likely to believe the false information. The four largest effects were based on political leaning, where, as compared to those on the left, those on the right are:
\begin{itemize}
    \setlength\itemsep{-0.5em}
    \item $11.87$ times more likely to believe that ``Gargling warm water with salt or vinegar eliminates infection from COVID-19", $95\%$ CI $[2.68, 52.55]$.
    \item $8.21$ times more likely to believe that ``House Democrats included $25$ million to boost their own salaries in their proposal for the COVID-19 related stimulus package", $95\%$ CI $[4.29, 15.72]$.
    \item $7.25$ times more likely to believe that ``Asymptomatic carriers of COVID-19 (infected but no symptoms) who die of other medical problems are added to the coronavirus death toll to get the numbers up to justify this pandemic response", $95\%$ CI $[3.92, 13.39]$.
    \item $6.30$ times more likely to believe that ``COVID-19 was man-made in a lab and is not thought to be a natural virus", $95\%$ CI $[3.25, 12.21]$.
\end{itemize}
The one false statement that those on the left are more likely to believe ($4.53$ times more likely, $95\%$ CI $[1.94, 10.54]$) than those on the right was, ``US President Donald Trump tweeted that economic stimulus checks would only go to those who have not made negative social media comments about him",

The effects of main news source on likelihood of believing false statements is smaller than the effects of political leaning. The two largest effects were that those with social media as their primary source are:
\begin{itemize}
    \setlength\itemsep{-0.5em}
    \item $3.14$ times more likely to believe that ``Drinking sodium bicarbonate and lemon juice reduces the acidity of the body and the risk of getting infected with COVID-19", $95\%$ CI $[1.49, 6.60]$.
    \item $2.97$ times more likely to believe that ``The COVID-19 outbreak is not actually caused by a virus, but by 5G-wireless phone technology", $95\%$ CI $[1.29, 6.83]$.
\end{itemize}
%
%


\subsection*{Western Europe (WE)}

On average, $42.7\%$~|~$24.0\%$ of true~|~false statements reached respondents, $43.4\%$~|~$11.7\%$ of true~|~false statements are believed, and $47.5\%$~|~$24.4\%$ of the true~|~false statements are believed by others known to the respondent. The median number of true~|~false statements that reached a respondent is $9$~|~$5$, the median number of true~|~false statements believed by a respondent is $9$~|~$2$, the median number of true~|~false statements believed by others known to the respondent is $10$~|~$4$, and $55.2\%$ believe at least one false conspiracy (see also~\cite{freeman2020}).

As before, we examined the effect of respondent demographics on the reach, belief, and belief by others in the true~|~false statements. Age is the only factor that had an effect on the likelihood of the number of false statements that are believed. The number of false statements believed by younger respondents\footnote{The younger and older age groups were created based on the median respondent age of $40$.} is $1.36$ times greater than by older respondents ($95\%$~CI [$1.15$, $1.60$]).

Examining the effect of political leaning and main news source on belief in each false statement revealed that political leaning influenced the likelihood of believing $3~|~21$ false statements and main news source influenced the likelihood of believing $2~|~21$ false statements. For $2~|~3$ false statements where there is an effect of political leaning those on the right are more likely to believe the false information. Specifically, as compared to those on the left, those on the right are:
\begin{itemize}
    \setlength\itemsep{-0.5em}
    \item $2.99$ times more likely to believe that ``COVID-19 was man-made in a lab and is not thought to be a natural virus", $95\%$ CI $[1.66, 5.37]$.
    \item $1.89$ times more likely to believe that ``Asymptomatic carriers of COVID-19 (infected but no symptoms) who die of other medical problems are added to the coronavirus death toll to get the numbers up to justify this pandemic response", $95\%$ CI $[1.10, 3.24]$.
\end{itemize}
The one false statement that those on the left are more likely to believe ($2.23$ times more likely, $95\%$ CI $[1.21, 4.09]$) than those on the right is that ``In a speech, President Vladimir Putin blamed the West for the spread of COVID-19".

For both false statements where there is an effect of main news source, those with social media as a main source are more likely to believe the false information. Specifically, those with social media as their primary source are:
\begin{itemize}
    \setlength\itemsep{-0.5em}
    \item $2.38$ times more likely to believe that ``Hand sanitizer has the same ingredients as antifreeze and can be fatal to pets if they lick your hand",$95\%$ CI $[1.36, 4.15]$.
    \item $2.22$ times more likely to believe that ``UK’s NHS banned alcohol consumption and sales in an effort to combat COVID-19", $95\%$ CI $[1.18, 4.17]$.
\end{itemize}
%
%


\subsection*{Vaccination}

Respondents’ intent to get vaccinated is significantly lower in the US ($59.3\%$) and WE ($64.0\%$) than in MENA ($74.6\%$) or CSA ($79.2\%$). For each region, we performed a binary logistic regression to examine whether respondent demographics are associated with intent to get vaccinated. In US, political leaning has the largest effect: those on the right are $3.17$ times less likely to say they will get vaccinated than those on the left, $95\%$ CI $[1.83, 5.49]$. In WE there are only small effects with the most substantial being that of education: those who have completed an undergraduate degree or higher are $1.94$ times more likely to say they will get vaccinated, $95\%$ CI $[1.32, 2.85]$. In CSA and MENA, there are no reliable effects of demographics on intent to get vaccinated.


\section*{Crowd Sourcing vs. Ipsos}

Because of the cost of recruiting a large representative sample, we wondered if standard crowd-sourcing surveys could replicate our basic findings. To this end, in addition to the Ipsos samples, a total of $601$, US-based respondents were recruited from Mechanical Turk (MTurk) workers. These respondents were an opportunity sample with no demographic quotas. The survey materials and procedure were identical to those described in the Methods section. All responses were collected between June 5, 2020 and July 14, 2020, amidst the global COVID-19 crisis. Based on responses to the attention-check questions, $100$ of the $601$ responses were discarded, yielding a total of $501$ usable responses. Respondents were paid \$3.00 for their participation in the study.

The pattern of responses for the two samples (MTurk and Ipsos) differ somewhat both in terms of reach and belief. On average, $54.6\%$~|~$35.8\%$ of true~|~false statements reached MTurk respondents (cf Ipsos: $42.3\%$~|~$27.1\%$), and $53.3\%$~|~$11.4\%$ of true~|~false statements are believed (cf Ipsos: $41.2\%$~|~$16.0\%$). Generally speaking, the MTurk respondents saw more COVID-related information, but were more likely to believe true statements and less likely to believe false statements. 

One possible explanation for these results is the demographic differences between the two samples. For example, $51\%$ of MTurk respondents identified as left of the political center, as compared to only $24\%$ of the Ipsos respondents; $52\%$ of MTurk respondents completed undergraduate studies as compared to $39\%$ of Ipsos respondents; and $16\%$ of the MTurk respondents were between $18-29$ years old, as compared to $26\%$ of the Ipsos respondents. 

To examine the impact of demographic differences, we matched the MTurk and Ipsos samples based on age, gender, education, and political affiliation. This yielded a total of $210$ matched MTurk samples. Even with this demographic matching, however, there was still a notable difference in reach and belief: $53.0\%$~|~$35.9\%$ of true~|~false statements reached MTurk respondents (cf Ipsos: $42.3\%$~|~$27.1\%$), and $51.6\%$~|~$13.7\%$ of true~|~false statements are believed (cf Ipsos: $41.2\%$~|~$16.0\%$). 

These results suggest that there might be a more fundamental difference between the two samples beyond simple demographics. If this is the case, then caution should be taken when using crowd sourcing to recruit a representative sample (see also, for example,~\cite{cooper2016Sun}).


\section*{Discussion}

We find, what we believe to be, a troubling global reach and belief in COVID-19 misinformation and conspiracies. Overall, the highest levels of reach and belief in false statements are in Central/South America (CSA) and the Middle East and North Africa (MENA). The reach of and belief in false statements is lower in Western Europe (WE) than in United States (US), yet true statements have similar reach and belief in these regions. This implies that the differences in WE and US are not one of access to false information, but the penetration of the false information. Belief in COVID-19 related misinformation is also more prevalent amongst respondents who consume news primarily on social media, and this association is more pronounced in the US than in the other three regions. Belief in misinformation is not partisan in CSA and MENA, is somewhat partisan in WE, and is extremely partisan in US.

The problem of misinformation seems to be particularly poignant in the United States where partisanship and news consumption on social media is leading to a particularly misinformed public. The consequences of an ill-informed public has real-world implications in the form of compliance with health recommendations and vaccination.


\section*{Acknowledgement}

This work was supported by a Seed Fund Award from CITRIS and the Banatao Institute at the University of California.




\pagebreak
\section*{Supplementary Materials}

\begin{small}
\begin{longtable}{lcc}
& Population demographics (\%) & Ipsos sample (\%)  \\ 
\hline
    \textbf{Gender}  &  &  \\
    \hline
    Man	    &	47.68	&	51.81	\\
    Woman	&	52.32	&	47.10	\\
    Prefer not to say	&	-	&	1.09	\\
    \hline
    
    \textbf{Age} &  &  \\
    \hline
    18 - 29	&	29.99	&	34.42	\\
    30 - 39	&	20.26	&	24.64	\\
    40 - 49	&	17.32	&	19.38	\\
    50 - 59	&	14.04	&	11.05	\\
    \textgreater{}60	&	18.39	&	3.62	\\
    Prefer not to say 	&	-	&	6.88	\\
    \hline

    \textbf{Geographic Location} & & \\
    \hline
    Argentina	&	11.19	&	9.60	\\
    Bolivia	&	2.87	&	0.72	\\
    Chile	&	4.72	&	9.42	\\
    Colombia	&	12.53	&	9.06	\\
    Costa Rica	&	1.26	&	1.81	\\
    Ecuador	&	4.33	&	8.88	\\
    El Salvador	&	1.61	&	4.53	\\
    Guatemala	&	4.13	&	4.17	\\
    Honduras	&	2.43	&	1.09	\\
    México	&	31.77	&	9.06	\\
    Nicaragua	&	1.63	&	1.81	\\
    Panamá	&	1.06	&	7.25	\\
    Paraguay	&	1.75	&	5.98	\\
    Perú	&	8.10	&	9.42	\\
    República Dominicana	&	2.67	&	1.63	\\
    Uruguay	&	0.86	&	6.70	\\
    Venezuela	&	7.10	&	8.88	\\
    \hline
    
    \textbf{Political Affiliation} & & \\
    \hline
    Strongly left-wing	&	-	&	3.80	\\
    Slightly left-wing	&	-	&	16.12	\\
    Center	&	-	&	58.88	\\
    Slightly right-wing	&	-	&	16.67	\\
    Strongly right-wing	&	-	&	4.53	\\
    \hline
\caption{Demographic details for Central and South America (CSA) From left to right, the values in second and third columns correspond to the population demographics (provided by Ipsos, where available) and sample demographics from our Ipsos study. Cells with '-' indicate unavailable data.}
\label{tab:datasetS2-CSA}
\end{longtable}
\end{small}

\newpage

\begin{small}
\begin{longtable}{lcc}
& Population demographics (\%) & Ipsos sample (\%)  \\ 
\hline
     \textbf{Gender}  &  & \\
    \hline
    Man	    &	53.67	&	60.08 \\
    Woman	&	46.33	&	37.98	\\
    Prefer not to say	&	-	&	1.94	\\
    \hline

    \textbf{Age} &  & \\
    \hline
    18 - 29	&	32.64	&	35.85	\\
    30 - 39	&	24.83	&	34.30	\\
    40 - 49	&	18.49	&	15.50	\\
    50 - 59	&	12.19	&	8.53	\\
    \textgreater{}60	&	11.85	&	1.74	\\
    Prefer not to say	&	-	&	4.07	\\
    \hline

    \textbf{Geographic Location} & & \\
    \hline
    Algeria	&	16.63	&	8.72	\\
    Egypt	&	38.78	&	11.82	\\
    Israel	&	3.50	&	20.93	\\
    Jordan	&	3.90	&	5.81	\\
    Kuwait	&	1.63	&	3.68	\\
    Lebanon	&	2.65	&	3.29	\\
    Morocco	&	14.09	&	11.43	\\
    Palestinian Territory	&	1.81	&	4.65	\\
    Saudi Arabia	&	13.24	&	19.38	\\
    United Arab Emirates	&	3.77	&	10.27	\\
    \hline
    
    \textbf{Political Affiliation} & & \\
    \hline
    Strongly left-wing	&	-	&	3.68	\\
    Slightly left-wing	&	-	&	7.95	\\
    Center	&	-	&	71.12	\\
    Slightly right-wing	&	-	&	12.02	\\
    Strongly right-wing	&	-	&	5.23	\\
    \hline
\caption{Demographic details for the Middle East and North Africa (MENA) From left to right, the values in second and third columns correspond to the population demographics (provided by Ipsos, where available) and sample demographics from our Ipsos study. Cells with '-' indicate unavailable data.}
\label{tab:datasetS2-MENA}
\end{longtable}
\end{small}

\newpage

\begin{small}
\begin{longtable}{lccc}
& Population demographics (\%) & Ipsos sample (\%) & MTurk sample (\%) \\ 
\hline
    \textbf{Gender}  &  &  &  \\
    \hline
    Man   & 48.30	&	50.58	&	46.91  \\
    Woman & 51.70	&	49.03	&	52.50  \\
    Prefer not to say & -	&	0.39	&	0.60  \\
    \hline
    
    \textbf{Age} &  &  &  \\
    \hline
    18 - 29 & 20.30	&	26.45	&	16.17  \\
    30 - 39 & 17.40	&	21.62	&	36.73  \\
    40 - 49 & 16.30	&	20.85	&	26.75  \\
    50 - 59 & 17.00	&	18.34	&	12.77  \\
    \textgreater{}60 & 29.00	&	10.04	&	6.59  \\
    Prefer not to say & -	&	2.70	&	1.00  \\ 
    \hline
    
    \textbf{Ethnicity} &  &  &  \\
    \hline
    American Indian + Alaska Native	 &  0.60	&	0.58	&	0.60 \\
    Black or African American &	11.70	&	10.81	&	7.19 \\
    Asian &	5.80	&	5.79	&	6.59 \\
    Hispanic &	16.30	&	8.30	&	3.79 \\
    Middle Eastern or North African &	1.00	&	0.19	&	0.60 \\
    Native Hawaiian or Other Pacific Islander &	0.20	&	0.58	&	0.00 \\
    White &	62.70	&	71.62	&	79.24 \\
    Other &	1.70	&	2.12	&	2.00  \\
    \hline

    \textbf{Education}  &   &   &  \\
    \hline
    None/little schooling completed  &  11.6	 &	2.32	&	0.00  \\
    Secondary education &  27.3	 &	34.17	&	24.35  \\
    Trade/vocational training & 8.4	&	11.00	&	12.38  \\
    Undergraduate &  41.2	&	38.61	&	51.90 \\
    Postgraduate &  11.5	&	13.90	&	11.18 \\
    \hline

    \textbf{Geographic Location} & & & \\
    \hline
    Alabama     &	1.50	&	2.12	&	1.80	\\
    Alaska	    & 0.20	&	0.00	&	0.00	\\
    Arizona 	&	2.20	&	1.35	&	2.79	\\
    Arkansas	&	0.90	&	0.97	&	0.40	\\
    California	&	12.10	&	6.18	&	9.18	\\
    Colorado	&	1.80	&	1.54	&	1.00	\\
    Connecticut	&	1.10	&	0.77	&	1.00	\\
    Delaware	&	0.30	&	0.19	&	0.00	\\
    District of Columbia	&	0.20	&	0.58	&	0.00	\\
    Florida	    &	6.80	&	9.07	&	8.58	\\
    Georgia	    &	3.20	&	6.95	&	4.59	\\
    Hawaii  	&	0.40	&	0.19	&	0.20	\\
    Idaho   	&	0.50	&	0.39	&	0.00	\\
    Illinois	&	3.90	&	4.25	&	3.79	\\
    Indiana	    &	2.00	&	0.97	&	1.20	\\
    Iowa    	&	0.90	&	0.77	&	0.60	\\
    Kansas	    &	0.90	&	2.12	&	1.20	\\
    Kentucky	&	1.40	&	0.97	&	2.20	\\
    Louisiana	&	1.40	&	1.35	&	0.80	\\
    Maine	    &	0.40	&	0.19	&	1.00	\\
    Maryland	&	1.90	&	1.35	&	1.40	\\
    Massachusetts	&	2.20	&	1.74	&	2.00	\\
    Michigan	&	3.10	&	3.28	&	4.59	\\
    Minnesota	&	1.70	&	1.74	&	1.00	\\
    Mississippi	&	0.90	&	0.39	&	0.80	\\
    Missouri	&	1.90	&	1.74	&	1.80	\\
    Montana	    &	0.30	&	0.39	&	0.20	\\
    Nebraska	&	0.60	&	0.39	&	0.40	\\
    Nevada	    &	0.90	&	0.39	&	1.60	\\
    New Hampshire	&	0.40	&	0.39	&	0.40	\\
    New Jersey	&	2.80	&	2.51	&	2.20	\\
    New Mexico	&	0.60	&	0.39	&	0.40	\\
    New York	&	6.10	&	8.49	&	6.59	\\
    North Carolina	&	3.20	&	3.67	&	3.39	\\
    North Dakota	&	0.20	&	0.39	&	0.00	\\
    Ohio	    &	3.60	&	3.28	&	4.79	\\
    Oklahoma	&	1.20	&	1.74	&	0.60	\\
    Oregon	    &	1.30	&	0.39	&	2.20	\\
    Pennsylvania	&	4.00	&	3.86	&	6.59	\\
    Rhode Island	&	0.30	&	0.39	&	0.60	\\
    South Carolina	&	1.60	&	1.16	&	0.80	\\
    South Dakota	&	0.30	&	0.39	&	0.40	\\
    Tennessee	&	2.10	&	3.47	&	1.20	\\
    Texas	    &	8.40	&	9.07	&	4.99	\\
    Utah	    &	0.90	&	0.39	&	0.40	\\
    Vermont	    &	0.20	&	0.19	&	0.20	\\
    Virginia	&	2.60	&	3.67	&	2.99	\\
    Washington	&	2.30	&	1.54	&	2.20	\\
    West Virginia	&	0.60	&	0.58	&	0.80	\\
    Wisconsin	&	1.80	&	1.74	&	2.00	\\
    Wyoming	    &	0.20	&	0.00	&	0.40	\\
    N/A	&	-	&	-	&	1.80	\\
    \hline

    \textbf{Political Affiliation} & & & \\
    \hline
    Strongly left-wing	&	-	&	12.36	&	25.35	\\
    Slightly left-wing	&	-	&	11.78	&	26.15	\\
    Center	            &	-	&	49.42	&	20.76	\\
    Slightly right-wing	&	-	&	13.90	&	17.96	\\
    Strongly right-wing	&	-	&	12.55	&	9.78	\\
    \hline
\caption{Demographic details for the United States (US) From left to right, the values in second through fourth columns correspond to the population demographics (provided by Ipsos, where available) and sample demographics from our Ipsos study, and sample demographcis from the MTurk study. Cells with '-' indicate unavailable data.}
\label{tab:datasetS2-US}
\end{longtable}
\end{small}

\newpage

\begin{small}
\begin{longtable}{lcc}
& Population demographics (\%) & Ipsos sample (\%)  \\ 
\hline
    \textbf{Gender}  &  &    \\
    \hline
    Man	    &	48.52	&	49.90 \\
    Woman	&	51.48	&	48.53 \\
    Prefer not to say	&	-	&	0.57 \\
    \hline

    \textbf{Age} &  &   \\
    \hline
    18 - 29	&	17.42	&	23.38	\\
    30 - 39	&	15.77	&	24.36	\\
    40 - 49	&	17.31	&	20.24	\\
    50 - 59	&	17.82	&	21.61	\\
    \textgreater{}60	&	31.67	&	8.64	\\
    Prefer not to say	&	-	&	1.77	\\
    \hline
    
    \textbf{Geographic Location} & & \\
    \hline
    Austria	&	2.30	&	8.25	\\
    Belgium	&	2.97	&	7.66	\\
    Denmark	&	1.50	&	9.43	\\
    France	&	17.34	&	9.43	\\
    Germany	&	21.49	&	9.04 \\
    Italy	&	15.59	&	9.04	\\
    Netherlands	&	4.48	&	9.23	\\
    Portugal	&	2.66	&	9.04	\\
    Spain	&	12.17	&	8.84 \\
    Switzerland	&	2.22	&	9.43	\\
    United Kingdom	&	17.28	&	10.61	\\
    \hline

    \textbf{Political Affiliation} & & \\
    \hline
    Strongly left-wing	&	-	&	7.66	\\
    Slightly left-wing	&	-	&	27.50	\\
    Center	&	-	&	42.83	\\
    Slightly right-wing	&	-	&	15.91	\\
    Strongly right-wing	&	-	&	6.09	\\
    \hline
\caption{Demographic details for Western Europe (WE) From left to right, the values in second and third columns correspond to the population demographics (provided by Ipsos, where available) and sample demographics from our Ipsos study. Cells with '-' indicate unavailable data.}
\label{tab:datasetS2-WE}
\end{longtable}
\end{small}


\end{document}